\documentclass[12pt,journal,onecolumn,draftcls,peerreview]{IEEEtran}

\usepackage{amsthm}
\usepackage{amssymb}
\usepackage{algorithm}
\usepackage{algorithmic}
\usepackage{amsmath}
\allowdisplaybreaks[4]
\usepackage{amsfonts}
\usepackage{graphicx}
\usepackage{makeidx}
\usepackage{subfigure}
\usepackage{cite}
\usepackage{epstopdf}

\IEEEoverridecommandlockouts

\graphicspath{{figures/}}

\begin{document}
%
% paper title
% Titles are generally capitalized except for words such as a, an, and, as,
% at, but, by, for, in, nor, of, on, or, the, to and up, which are usually
% not capitalized unless they are the first or last word of the title.
% Linebreaks \\ can be used within to get better formatting as desired.
% Do not put math or special symbols in the title.
\title{A New Hybrid Half-Duplex/Full-Duplex Relaying System with Antenna Diversity}

\author{
\IEEEauthorblockN{Cheng Li, ~Bin Xia, ~Zhiyong Chen %Shihai Shao\IEEEauthorrefmark{2}, Youxi Tang\IEEEauthorrefmark{2}
}

\IEEEauthorblockA{Department of Electronic Engineering, Shanghai Jiao Tong University, Shanghai, China}

%\IEEEauthorblockA{\IEEEauthorrefmark{2} University of Electronic Science and Technology of China, Chengdu, China}

Emails: \{lichengg, bxia, zhiyongchen\}@sjtu.edu.cn%,  ~\IEEEauthorrefmark{2}\{ssh, tangyx\}@uestc.edu.cn

% note the % following the last \IEEEmembership and also \thanks -
% these prevent an unwanted space from occurring between the last author name
% and the end of the author line. i.e., if you had this:
%
% \author{....lastname \thanks{...} \thanks{...} }
%                     ^------------^------------^----Do not want these spaces!
%
% a space would be appended to the last name and could cause every name on that
% line to be shifted left slightly. This is one of those "LaTeX things". For
% instance, "\textbf{A} \textbf{B}" will typeset as "A B" not "AB". To get
% "AB" then you have to do: "\textbf{A}\textbf{B}"
 \thanks {This work was supported in part by the National Key Science and Technology Specific Program under Grant 2016ZX03001015, in part by the National Nature Science Foundation of China under Grant 61531009, and in part by the National High Technology Research and Development Program of China under 863 5G Grant 2014AA01A704.}
 }
% that ends a line with a % and do not let a space in before the next \thanks.
% Spaces after \IEEEmembership other than the last one are OK (and needed) as
% you are supposed to have spaces between the names. For what it is worth,
% this is a minor point as most people would not even notice if the said evil
% space somehow managed to creep in.

\maketitle
\begin{abstract}
The hybrid half-duplex/full-duplex (HD/FD) relaying scheme is an effective paradigm to overcome the negative effects of the self-interference incurred by the full-duplex (FD) mode. However, traditional hybrid HD/FD scheme does not consider the diversity gain incurred by the multiple antennas of the FD node when the system works in the HD mode, leading to the waste of the system resources. In this paper, we propose a new hybrid HD/FD relaying scheme, which utilizes both the antennas of the FD relay node for reception and transmission when the system works in the HD mode. With multiple antennas, the maximum ratio combining/maximum ratio transmission is adopted to process the signals at the relay node. Based on this scheme, we derive the exact closed-form system outage probability and conduct various numerical simulations. The results show that the proposed scheme remarkably improves the system outage performance over the traditional scheme, and demonstrate that the proposed scheme can more effectively alleviate the adverse effects of the residual self-interference. %The results show that the system can reap the diversity gain with the proposed scheme, and demonstrate the superiority of the proposed scheme over the traditional scheme.
\end{abstract}
% Note that keywords are not normally used for peerreview papers.
%\begin{IEEEkeywords}
%Instantaneous Outage Probability, Two-way Relaying, Full-duplex, Nakagami-m fading, Decode-and-Forward, \LaTeX.
%\end{IEEEkeywords}

% For peer review papers, you can put extra information on the cover
% page as needed:
% \ifCLASSOPTIONpeerreview
% \begin{center} \bfseries EDICS Category: 3-BBND \end{center}
% \fi
%
% For peerreview papers, this IEEEtran command inserts a page break and
% creates the second title. It will be ignored for other modes.
%\IEEEpeerreviewmaketitle

%\vspace {1em}
\section{Introduction}
The full-duplex (FD) communications are receiving more and more interest from both industry and academia due to the ultimate utilization of radio resources \cite{7105650,7194828}. Compared to the half-duplex (HD) mode, the FD mode bears the capability of reception and transmission on the same time and frequency resource \cite{7224732}. The FD mode is usually achieved with two antennas, one for reception and the other for transmission \cite{6353396}. The signals leakage, which is called self-interference, from the transmit antenna to the local receive antenna would severely limit the system performance. Although the self-interference can be greatly cancelled with various methods \cite{5985554}, the performance gain of the FD mode over the HD mode is limited by the residual self-interference (RSI).

On the other side, the relay stations are usually deployed in the remote area to extend the coverage of the cellular networks. Integrating the FD mode into the relay communication is an effective way to improve the rate of the cell edge users \cite{1341264}. The performance of the FD relay system have been investigated in \cite{7410116,6340386,7801128,6832455,6812188,5466077}. Specifically, in \cite{7410116,6340386,7801128}, the authors have analyzed the achievable rate and system outage probabilities of the FD relay networks. The diversity gain of the FD relay system with direct link is investigated by the authors in \cite{6832455}. In addition to the performance analysis of the FD relay systems, the research has also shown that the FD system could achieve better performance compared to their HD counterparts when the RSI is below a certain threshold\cite{6812188, 5466077}.

To alleviate the adverse effects of the RSI, the hybrid half-duplex/full-duplex (HD/FD) relaying scheme was proposed \cite{5961159}. With the hybrid HD/FD relaying scheme, the system can dynamically change between the HD mode and FD mode. When the RSI is high, the system converts to the HD mode, thus, the RSI can be inherently eliminated. However, in the previous works, although the authors considered the hybrid HD/FD relaying scheme, the system resources have not been fully utilized. Specifically, the authors only used one antenna to receive and transmit signals when the system works in the HD mode even if the FD relay node is equipped with two antennas\cite{5961159}. This leads to certain diversity loss, resources waste and performance degradation.

In this paper, we propose a new hybrid HD/FD relaying scheme. In this scheme, the MRC/MRT, which utilizes the antenna diversity to combat the channel fading, is adopted to process signals at the FD relay node when the system works in the HD mode. For performance evaluation, we derive the exact closed-form system outage probability, which is based on the FD outage probability and the conditional HD outage probability of the proposed HD/FD relaying scheme. In addition, we compare the proposed scheme with the traditional scheme proposed in \cite{5961159} in terms of the outage performance. The results show that the proposed scheme improves the system outage performance over the traditional scheme and alleviate the adverse effects of the RSI. Finally, numerical simulations corroborates the theoretical results.

%In this scheme, when the system works in the HD mode, the relay node utilizes both the antennas of the FD node to receive and transmit signals. The maximum ratio combining/maximum ratio transmission (MRC/MRT) is adopted to process the signals.To evaluate how much performance gain can be achieved of the proposed scheme, we consider the outage probability as the performance metric and the exact closed-form system outage probability is derived. In addition, we compare the outage probability of the proposed scheme with the traditional scheme via numerical simulations. The results demonstrate that the proposed hybrid HD/FD relaying scheme dramatically improves the outage performance over the traditional scheme, and show that proposed scheme efficiently alleviates the adverse effects of the RSI. %The Monte Carlo simulations guarantee the correctness of the analytical results.

%\newpage
%\mbox{}
%\newpage
\section{System Model}
In this section, we elaborate on the channel mode, the proposed hybrid HD/FD relaying scheme and the specific signaling process.

\begin{figure}[ht]
  \centering
  % Requires \usepackage{graphicx}
  \includegraphics[width=3.6 in]{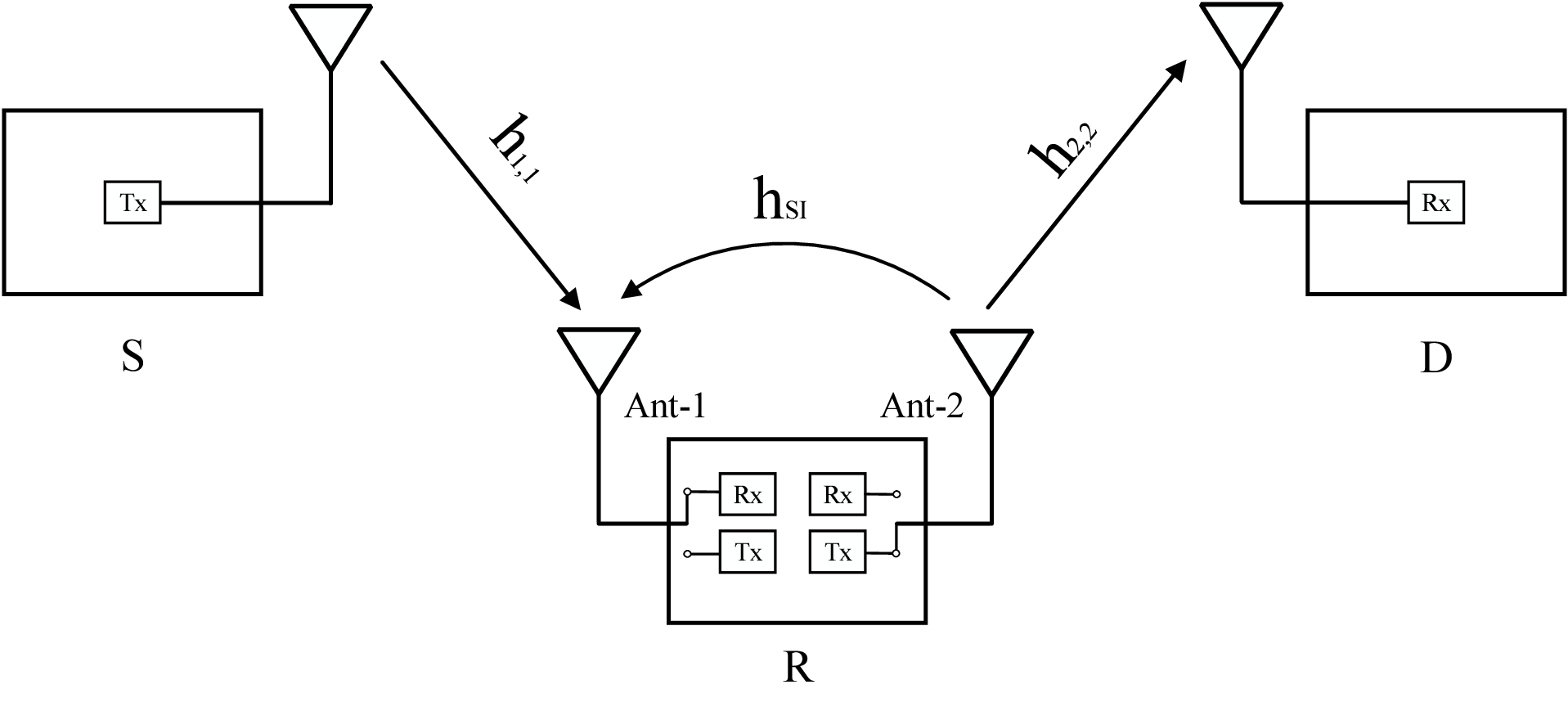}\\
  \vspace{0mm}
  \caption{The hybrid HD/FD relay system with FD mode.}\label{system_model}
\end{figure}

\begin{figure}[ht]
\centering
\subfigure[Sub-time slot 1.]{
\begin{minipage}{8cm}
\begin{centering}
    \includegraphics[width=3.3 in]{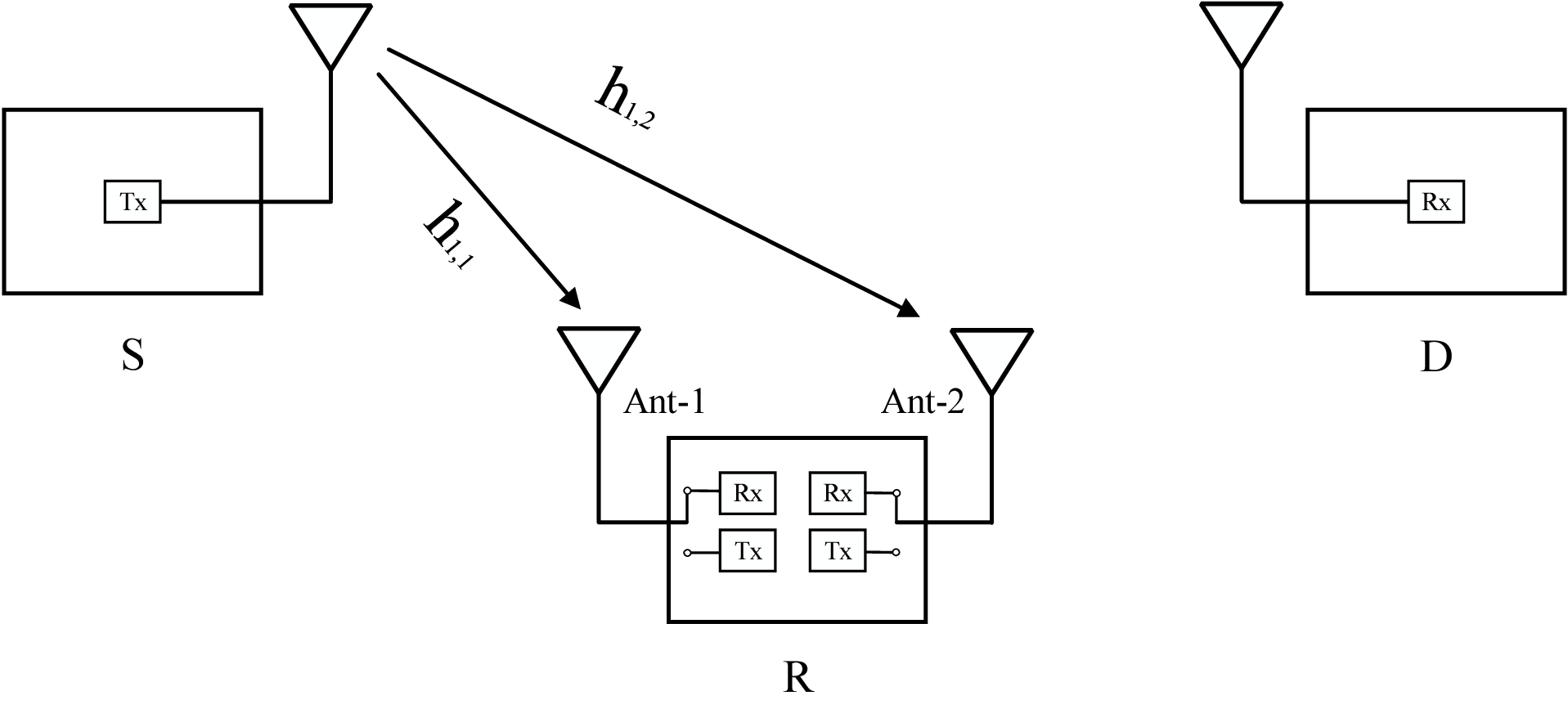}\\
    \end{centering}
    \hskip 0.5mm
    \end{minipage}
    }

    \subfigure[Sub-time slot 2.]{
    \begin{minipage}{8cm}
    \centering
    \includegraphics[width=3.3 in]{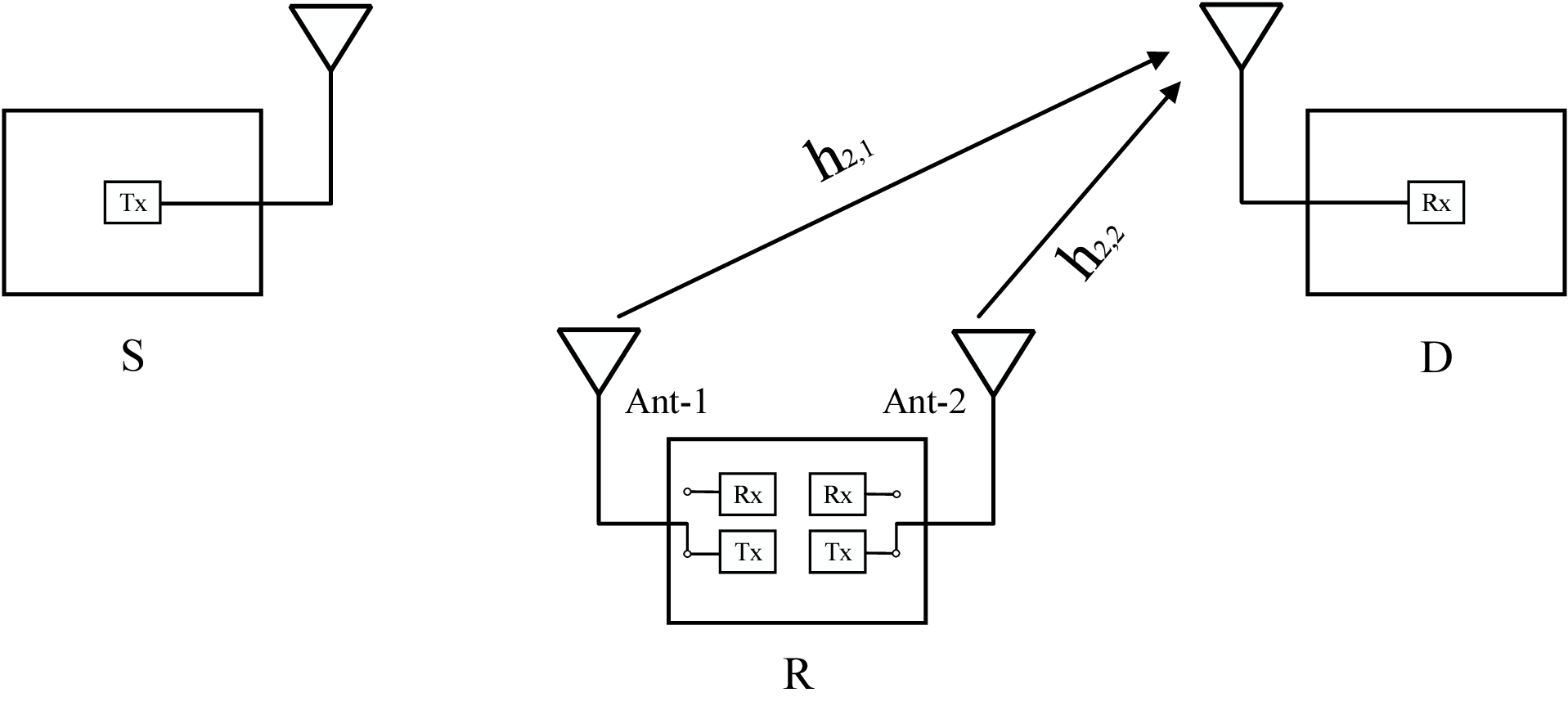}\\
   \hskip 0.2mm
    \end{minipage}
    }
    \vspace{-2mm}
    \caption{The hybrid HD/FD relay system with HD mode.}
\end{figure}

\subsection{Channel Model}
In this paper, we consider the one-way decode-and-forward FD relay system, which consists of a source node S, a destination node D and a relay node R. The source node S and the destination node D are both equipped with one antenna and can only work in the HD mode. However, the relay node R is equipped with two antennas Ant-1 and Ant-2, and each antenna is equipped with a Rx module and a Tx module. Thus, the relay node has the capability to work in either the FD mode or the HD mode. We assume that the direct link between the source node S and the destination node D does not exist due to the deep fading and strong shadowing effects \cite{6812188}. %However, the source node S can transmit messages to the destination node D via the relay node R.

%We use $h_{1,1}$, $h_{1,2}$ to denote the channel coefficients from the source node S to the antennas Ant-1 and Ant-2 of the relay node, respectively, and $h_{2,1}$, $h_{2,2}$ to denote the channel coefficients from the antennas Ant-1 and Ant-2 of the relay node to the destination node D, respectively.
The channel coefficients of $h_{1,1}$, $h_{1,2}$ and $h_{2,1}$, $h_{2,2}$ are shown in the Fig. 1. Since the antennas Ant-1 and Ant-2 can naturally achieve 40 dB isolation, thus $h_{1,1}$ and $h_{1,2}$, $h_{2,1}$ and $h_{2,2}$ are assumed to be independent with each other \cite{5961159}.  We model the links from the source S to the relay R and the relay R to the destination D as complex Gaussian channels. Then, we can easily prove that the envelops of the channels are subject to the Rayleigh distribution. Let $E\{|h_{i,j}|^2\}=\Omega_{i,j},\, i,\, j \in\{1, 2\}$ to denote the channel parameters. In this work, we assume that all the links are block fading wireless channels. %In addition, we assume that the transmit powers of the source node and the relay node are denoted by $P_{S}$ and $P_{R}$, respectively.

\subsection{The Proposed Hybrid HD/FD Relaying Scheme}
The proposed hybrid HD/FD scheme is elaborated on in the following.

\textbf{The Full-Duplex Mode:} The source node S transmits signals to the relay node, and the relay node R forwards the previously decoded signals to the destination node D simultaneously as depicted in Fig. 1. In the FD mode, we assume that the antenna Ant-1 is connected to the Rx module to receive signals from the source node S and the antenna Ant-2 is connected to the Tx module to forward the signals to the destination node D. Hence, the simultaneous reception and transmission is achieved. It is worth noting that in the FD mode, although several self-interference cancellation techniques can be adopted to cancel the self-interference, the system is harassed by the RSI \cite{5961159}.

\textbf{The Half-Duplex Mode:} The source node S transmits signals to the relay node, and the relay node R forwards the decoded signals to the destination node D in the next time slot as depicted in Fig.  2. In the HD mode, the system is divided into two phases: reception phase and relaying phase. During the reception phase, both the two antennas Ant-1 and Ant-2 of the relay node are connected to the Rx modules to receive signals from the source node S. Within the relaying phase, both the two antennas Ant-1 and Ant-2 are connected to the Tx modules to forward the recoded signals to the destination node D. In order to reap the diversity gain, we adopt the maximum ratio combining (MRC) to combine the signals received at the two antennas and the maximum ratio transmission (MRT) to forward the signals to the node D.

\textbf{The Hybrid Scheme:} In our proposed hybrid HD/FD scheme, the system works either in the FD mode or the HD mode determining on the instantaneous channel capacity.  If the instantaneous capacity of the FD mode surpasses that of the HD mode, i.e., $C_{fd}>C_{hd}$, the system chooses to work in the FD mode. The benefits of the FD mode are two-fold: i) The FD mode could inherently achieve higher spectrum efficiency; ii) The latency at the relay node could be greatly reduced. On the other side, if the instantaneous capacity of the HD mode is larger than that of the FD mode, i.e., $C_{hd}>C_{fd}$, the system converts to the HD mode. The benefits of the HD mode are also two-fold: i) The system can inherently eliminate the self-interference perfectly; ii) The MRC/MRT can be adopted to combat the channel fading.

%In our proposed hybrid HD/FD scheme, with the given minimum data threshold $R_{0}$ the system firstly works in the FD mode if the data could successfully delivered from the source node S to the destination node D. The benefits of the FD transmission is two-fold: i) The FD mode could inherently achieve higher spectrum efficiency; ii) The latency at the relay node is only the processing delay $\tau$, which is far shorter than the time slot. If the messages cannot be successfully transmitted from the source S to the destination R due to the severe residual self-interference or the deep channel fading, the relay node convert to the HD mode. With the HD mode, one time slot is divided into two sub-time slot. In the first sub-time slot, the relay node receives the signals from the source node S with two-antennas using the MRC combining. In the second sub-time slot, the relay node forwards signals to the destination node with both antennas using the MRT technique.  The benefits of the HD transmission is also two-fold: i) The system could inherently eliminate the self-interference perfectly; ii) The MRC/MRT can be adopted to combat the channel fading.

\subsection{The Specific Signaling Process}
\textbf{The FD mode}: In this mode, the signals received at the relay node and the destination node can be expressed as
\begin{equation}
y_{r}=h_{1,1}x_{s}+h_{si}x_{r}+n_{r},
\end{equation}
\begin{equation}
y_{d}=h_{2,2}x_{r}+n_{d},
\end{equation}
respectively, where $x_{s}$ and $x_{r}$ are the signals transmitted by the source S and the relay R with power $P_{S}=E\{|x_{s}|^2\}$ and $P_{R}=E\{|x_{r}|^2\}$, respectively. $n_{r}$ and $n_{d}$ denote the thermal noise over the relay R and the destination D, respectively, with zero mean and variance $\sigma^2$, i.e., $n_{r}, n_{d} \backsim \mathcal{CN}(0, \sigma^2)$. The relay R receives the signals from the source S as well as the self-interference signals from the antenna Ant-2 to Ant-1. $h_{si}$ denotes the self-interference channel coefficient. In this paper, we assume that the relay node could apply the self-interference cancellation techniques in \cite{5985554} to cancel $h_{si}x_{r}$. Hence, the system is harassed by the RSI after self-interference cancellation. We denote the RSI as $\widetilde{h}_{si}\widetilde{x}_{r}$. Whatever the specific distributions of $\widetilde{h}_{si}$ and $\widetilde{x}_{r}$ are, due to the imperfect estimation of the self-interference channel and distortion of the self-interference signals during the cancellation process, we assume that the effects of $\widetilde{h}_{si}\widetilde{x}_{r}$ are characterized by the Gaussian distribution \cite{5961159}, i.e., $\widetilde{h}_{si}\widetilde{x}_{r} \backsim\mathcal{CN}(0,\sigma_{RSI}^{2})$, where $\sigma_{RSI}^{2}=K_{r}P_{R}$. $K_{r}$ indicates the self-interference capability of the relay node. Thus, the received signals at the relay node after self-interference cancellation can be expressed as
\begin{equation}
\widetilde{y}_{r}=h_{1,1}x_{s}+\widetilde{h}_{si}\widetilde{x}_{r}+n_{r},
\end{equation}

Hence, the signal-to-interference-plus-noise ratio (SINR) at the relay node and the destination node can be expressed as
\begin{equation}
  \gamma_{f,r}=\frac{|h_{1,1}|^2 P_{S}}{k_{r}P_{R}+\sigma^2}, \quad \gamma_{f,d}=\frac{|h_{2,2}|^2 P_{R}}{\sigma^2},
\end{equation}
where $\gamma_{f,r}$ and $\gamma_{f,d}$ denote the SINRs at the relay node R and the destination node D under the FD mode, respectively.

\textbf{The HD Mode:} In this mode, the time slot is divided into two sub-time slots. In the first sub-time slot, the relay node R receives signals from the source node S
\begin{equation}
\mathbf{y}_{r}=\mathbf{H}_{1}x_{s}+\mathbf{n}_{r}
\end{equation}
where vector $\mathbf{y}_{r}=[y_{1,r}, y_{2,r}]^{T}$ denotes the received signals at the antennas Ant-1 and Ant-2. $\mathbf{H}_{1}=[h_{1,1},h_{1,2}]^{T}$ denotes the estimated channel vector between the source S and the relay R. $\mathbf{n}_{r}=[n_{1,r}, n_{2,r}]^{T}$ denotes the Gaussian noises over the antennas Ant-1 and Ant-2 and we assume $n_{1,r}, n_{2,r}\sim \mathcal{CN}(0, \sigma^2)$. In order to maximize the SINR at the relay R, we adopt the MRC to combine the received signals at the antennas Ant-1 and Ant-2. The combined signals can be expressed as

\begin{equation}
\mathbf{y'}_{r}=\mathbf{W_{1}}^{H}\mathbf{H}_{1}x_{s}+\mathbf{W_{1}}^{H}\mathbf{n}_{r}
\end{equation}
where $(\cdot)^H$ denotes the conjugate transpose, $\mathbf{W_{1}}=\frac{\mathbf{H}_{1}}{||\mathbf{H}_{1}||_{F}}$ denotes the processing matrix of the MRC and $||\cdot||_{F}$ denotes the Frobenius norm.

In the second sub-time slot, the relay node R uses the MRT technique to pre-process the signals and then forwards the signals to the destination node D. The received signals at the destination node can be expressed as
\begin{equation}
y_{d}=\mathbf{W_{2}}^{H}\mathbf{H_{2}}x_{r}+n_{d},
\end{equation}
where $\mathbf{W_{2}}=\frac{\mathbf{H}_{2}}{||\mathbf{H}_{2}||_{F}}$ is the processing matrix of the MRT and $\mathbf{H}_{2}=[h_{2,1}, h_{2,2}]^T$ is the estimated channel vector from the antennas at the relay node R to the destination node D.

Based on the MRC/MRT, the SINR at the relay node R and the destination node D can be expressed as
\begin{equation}
  \gamma_{h,r}=\frac{|h_{1,1}|^2+|h_{1,2}|^2}{\sigma^2}P_{S},
\end{equation}
\begin{equation}
  \gamma_{h,d}=\frac{|h_{2,1}|^2+|h_{2,2}|^2}{\sigma^2}P_{R}
\end{equation}
where $ \gamma_{h,r}$ and $\gamma_{h,d}$ denote the SINRs at the relay node and the destination node under the HD mode, respectively.

\section{Outage Performance Analysis}
In this section, we analyze the system outage performance of the considered one-way decode-and-forward FD relay system under the proposed hybrid HD/FD relaying scheme.

The system outage probability under the FD mode and HD mode can be defined as
\begin{equation}
  P_{out}^{fd} = Pr\{C_{fd}=\log_{2}(1+\min\{\gamma_{f,r}, \gamma_{f,d}\})<R_{0}\},
\end{equation}
\begin{equation}
  P_{out}^{hd} = Pr\{C_{hd}=\frac{1}{2}\log_{2}(1+\min\{\gamma_{h,r}, \gamma_{h,d}\})<R_{0}\},
\end{equation}
respectively, where $C_{fd}$ and $C_{hd}$ denote the system capacity of the FD mode and the HD mode, respectively. $Pr\{x\}$ denotes the probability of the event $x$. We can note that due to the simultaneous transmission, the pre-factor 1/2 disappears under the FD mode, which indicates that the FD mode could effectively recover the spectrum efficiency.

With the proposed hybrid HD/FD scheme, the system outage probability can be calculated as
\begin{align}
P_{out}^{sys}= &\ Pr\{C_{fd}<R_{0}, C_{hd}<R_{0}\}\notag\\
= &\ Pr\{C_{fd}<R_{0}\}Pr\{C_{hd}<R_{0}|\, C_{fd}<R_{0}\},
\end{align}
%which can be explained as that the relay node R firstly works in the FD mode. Unfortunately, according to the channel estimation results, the system control node\footnote{In this paper, the relay node is assumed to be the control center that performs the mode decision.} finds that the messages cannot be successfully delivered with the minimum required rate $R_{0}$ from the source node to the destination node. Then, the relay node converts to the HD mode under the condition of the FD system outage. If the relay node utilizes two antennas Ant-1 and Ant-2 under the HD mode, the messages still cannot be sustained with the minimum rate $R_{0}$, the system is defined in the outage state.

According to the Total Probability Theorem, the system outage probability of the FD mode can be divided into three mutual exclusive events A, B and C as follows
\begin{eqnarray}
P_{out}^{fd} &=&Pr\{\underbrace{C_{fd}^{sr}<R_{0},\ C_{fd}^{rd}>R_{0}}_{A}\}\notag\\
&+&Pr\{\underbrace{C_{fd}^{sr}>R_{0}, \ C_{fd}^{rd}<R_{0}}_{B}\}\notag\\
&+&Pr\{\underbrace{C_{fd}^{sr}<R_{0},\ C_{fd}^{rd}<R_{0}}_{C}\},
\end{eqnarray}
where $C_{fd}^{sr}=\log_{2}(1+\gamma_{f,r})$ and $C_{fd}^{rd}=\log_{2}(1+\gamma_{f,d})$ denote the channel capacities of the links from the source S to the relay R and from the relay R to the destination D, respectively. Event $A$ denotes that the link from the source S to the relay R is in the outage state but the link from the relay R to the destination D is not. Event $B$ denotes that the link from the source S to the relay R is not in the outage state whereas the link from the relay R to the destination D is. Event $C$ denotes that both the links from the source S to the relay R and from the relay R to the destination D are in the outage state.

Applying the Total Probability Theorem again, the system outage probability can be expressed as
\begin{eqnarray}\label{sys-out}
P_{out}^{sys} &=& Pr\{A\}Pr\{C_{hd}<R_{0}|A\}\notag\\
&+&Pr\{B\}Pr\{C_{hd}<R_{0}|B\}\notag\\
&+&Pr\{C\}Pr\{C_{hd}<R_{0}|C\}.
\end{eqnarray}

Next, we will derive the system outage probabilities under the conditions of events A, B and C, respectively.

\subsection{System Outage Probability Under the Event A}

Recall that the complex Gaussian channels have the parameter $E\{|h_{i,j}|^2\}=\Omega_{i,j}, i, j \in\{1, 2\}$, we can easily prove that the random variable $|h_{i,j}|^2$ is subject to the exponential distribution. The probability density function (pdf) of $|h_{i,j}|^2$ is expressed as
\begin{equation}
f_{|h_{i,j}|^2}(x) = \lambda_{i,j}e^{(-\lambda_{i,j}x)},
\end{equation}
where $\lambda_{i,j}=\frac{1}{\Omega_{i,j}},\, i,\, j \in\{1,2\}$.

Thus, we have
\begin{eqnarray}\label{Prob-A}
Pr\{A\}\!\!\! &=&\!\!\! Pr\{C_{fd}^{sr}<R_{0},\ C_{fd}^{rd}>R_{0}\} \notag\\
            &=&\!\!\! Pr\{\log_{2}(1+\gamma_{f,r})<R_{0}, \log_{2}(1+\gamma_{f,d})>R_{0}\} \notag\\
            &=&\!\!\! Pr\{\frac{|h_{1,1}|^2 P_{S}}{K_{R}P_{R}+\sigma^2} < t_{1}, \frac{|h_{2,2}|^2 P_{R}}{\sigma^2}>t_{1}\}\notag\\
            &=&\!\!\!\int_{0}^{\frac{t_{1}(K_{R}P_{R}+\sigma^2)}{P_{S}}}\!\!\int_{\frac{t_{1}\sigma^2}{P_{R}}}^{+\infty}f_{|h_{1,1}|^2}(x)f_{|h_{2,2}|^2}(y)dxdy \notag\\
            &=&\!\!\!(1-e^{(-\lambda_{1,1}t_{1}\frac{K_{R}P_{R}+\sigma^2}{P_{S}})})e^{(-\lambda_{2,2}\frac{t_{1}\sigma^2}{P_{R}})},
\end{eqnarray}
where $t_{1}=2^{R_{0}}-1$. Under the condition of the event A, the outage probability of the HD mode can be calculated as
\begin{eqnarray}\label{out-A}
Pr\{C_{hd}<R_{0}|A\}\!\!\!&=& \!\!\! Pr\{C_{hd}^{sr}<R_{0}|A\}\notag\\
&+&\!\!\!Pr\{C_{hd}^{rd}<R_{0}|A\}\notag \\
&-&\!\!\!Pr\{C_{hd}^{sr}<R_{0}, C_{hd}^{rd}<R_{0}|A\},
\end{eqnarray}
where $C_{hd}^{sr}$ and $C_{hd}^{rd}$ denote the channel capacity of the links from the source S to the relay R and from the relay R to the destination D, respectively.

Then, we first calculate the $Pr\{C_{hd}^{sr}<R_{0}|A\}$
\begin{small}
\begin{align}\label{out-A-1}
 & Pr\{C_{hd}^{sr}<R_{0}|A\} \notag\\
 =&Pr\{\frac{1}{2}\log_{2}(1+\frac{|h_{1,1}|^2+|h_{1,2}|^2}{\sigma^2}P_{S})<R_{0}|A\}\notag\\
=&Pr\{\frac{(|h_{1,1}|^2 +|h_{1,2}|^2)P_{S}}{\sigma^2} <t_{2}\big||h_{1,1}|^2<\frac{t_{1}(K_{R}P_{R}+\sigma^2)}{P_{S}} \}\notag\\
=&\int_{0}^{\min\{\frac{t_{1}(K_{R}P_{R}+\sigma^2)}{P_{S}}, \frac{t_{2}\sigma^2}{P_{S}}\}}\!\!\!\!\!\int_{0}^{\frac{t_{2}\sigma^2}{P_{S}}-x}\!\!\!\!\! \!\!\!\!\! f_{(|h_{1,1}|^2|A)}(x)f_{|h_{1,2}|^2}(y)dxdy\notag \\
=&\int_{0}^{\min\{m_{1},m_{2}\}}\int_{0}^{\frac{t_{2}\sigma^2}{P_{S}}-x}\frac{f_{|h_{1,1}|^2(x)}}{P_{|h_{1,1}|^2}(m_{1})}f_{|h_{1,2}|^2}(y)dxdy \notag\\
=&\frac{1}{P_{|h_{1,1}|^2}^A(m_{1})}\big\{(1-e^{(-\lambda_{1,1}\min\{m_{1},m_{2}\})})\notag\\
-&\frac{\lambda_{1,1}e^{(-\lambda_{1,1}m_{2})}}{\lambda_{1,2}-\lambda_{1,1}}(e^{((\lambda_{1,2}-\lambda_{1,1})\min\{m_{1},m_{2}\})}-1)\big\},
\end{align}
\end{small}
where $t_{2}=4^{R_{0}}-1$, $m_{1}=\frac{t_{1}(K_{R}P_{R}+\sigma^2)}{P_{S}}$ and $m_{2}=\frac{t_{2}\sigma^2}{P_{S}}$. $P_{|h_{1,1}|^2}^{A}(m_{1})=1-e^{-\lambda_{1,1}m_{1}}$ denotes the probability of $|h_{1,1}|^2<m_{1}$ under event A.

By the similar way, we can derive the probability of $Pr\{C_{hd}^{rd}<R_{0}|A\}$ as follows
\begin{align}\label{out-A-2}
  &Pr\{C_{hd}^{rd}<R_{0}|A\} \notag\\
=&\int_{0}^{m_{3}}\int_{0}^{m_{3}-y}f_{(|h_{2,2}|^2|A)}(y)f_{(|h_{2,1}|^2)}(x)dxdy\notag\\
=&\frac{1}{P_{|h_{2,2}|^2}^{A}(m_{3})}\big\{(e^{(-\lambda_{2,2}m_{3})}-e^{(-\lambda_{2,2}m_{2}))}\notag\\
-&\frac{\lambda_{2,2}e^{(-\lambda_{2,1}m_{2})}}{\lambda_{2,1}-\lambda_{2,2}}(e^{((\lambda_{2,1}-\lambda_{2,2})m_{2})}-e^{((\lambda_{2,2}-\lambda_{2,1})m_{3})})\big\},
\end{align}
where $m_{3}=\frac{t_{1}\sigma^2}{P_{R}}$. $P_{|h_{2,2}|^2}^{A}(m_{3})=e^{-\lambda_{2,2}m_{3}}$  denotes the probability of $|h_{2,2}|^2>m_{3}$ under event A.

Next, we derive the probability of both the links from the source S to the relay R and from the relay R to the destination D locate in the outage region under the HD mode, i.e.,
\begin{align}\label{out-A-3}
&Pr\{C_{hd}^{sr}<R_{0}, C_{hd}^{rd}<R_{0}|A\}\notag\\
=&\frac{1}{P(A)}\!\!\!\int_{0}^{m_{\min}}\!\!\!\!\!\int_{0}^{m_{2}-y_{1}}\!\!\!\!\!\int_{ m_{3}}^{m_{2}}\!\!\!\!\!\int_{0}^{m_{2}-y_{2}}\!\!\!\!\!\!\!\!\!\!\!\!\!\!\!\!\!\!f(x_{1})f(y_{1})f(x_{2})f(y_{2})dx_{1}dy_{1}dx_{2}dy_{2} \notag\\
\overset{(a)}=&Pr\{C_{hd}^{sr}<R_{0}|A\}\times Pr\{C_{hd}^{rd}<R_{0}|A\},
\end{align}
where $m_{\min}=\min\{m_{1},m_{2}\}$. $P(A)=P_{|h_{1,1}|^2}^A(m_{1})\times P_{|h_{2,2}|^2}^A(m_{2})$ denotes the probability of the whole event A. (a) is achieved by the independence of the links from the source S to the relay R and from the relay R to the destination D. $f(x_{1})=f_{|h_{1,1}|^2}(x)$, $f(y_{1})=f_{|h_{1,2}|^2}(y_{1})$, $f(x_{2})=f_{|h_{2,1}|^2}(x_{2})$ and $f(y_{2})=f_{|h_{2,2}|^2}(y_{2})$ are the pdfs of the corresponding channels.

Substituting (\ref{out-A-1}), (\ref{out-A-2}) and (\ref{out-A-3}) into (\ref{out-A}), We can get the system outage probability under the condition of event A. Similar to the event A, the system outage probability under the event B and event C can also be derived in the similar methods. Due to the length limit of this paper, the details of the specific derivations procedures are omitted here. Then according to the (\ref{sys-out}),  we can obtain the whole system outage probability of the proposed hybrid HD/FD relaying scheme.

\section{Simulation Results}
In this section, we investigate the proposed hybrid HD/FD relaying scheme by the numerical simulations.
\begin{figure}[ht]
  \centering
  % Requires \usepackage{graphicx}
  \includegraphics[width=3.6 in]{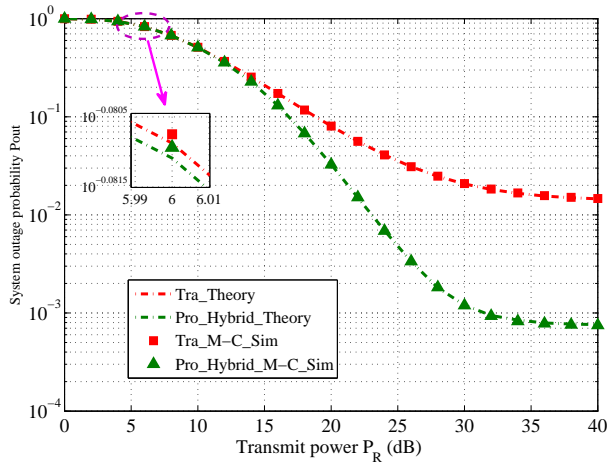}\\
  \vspace{0mm}
  \caption{The comparison with regard to the transmit power $P_{R}$.}\label{examp2}
\end{figure}

\begin{figure}[ht]
  \centering
  % Requires \usepackage{graphicx}
  \includegraphics[width=3.6 in]{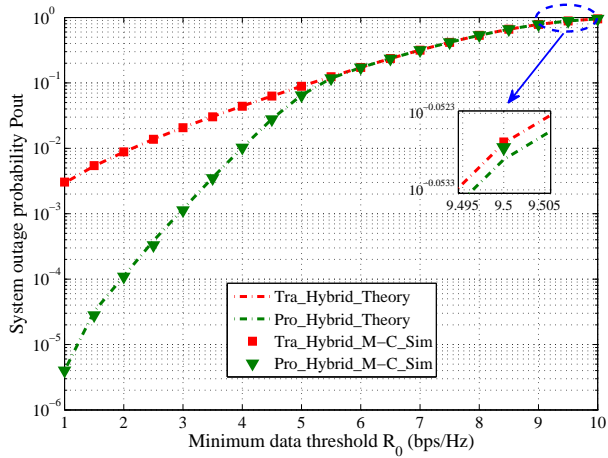}\\
  \vspace{0mm}
  \caption{The comparison with regard to the minimum data rate $R_{0}$ .}\label{examp3}
\end{figure}

In Fig. 3, we compare the traditional HD/FD relaying scheme in \cite{5961159} with the proposed HD/FD relaying scheme with the variation of the transmit power $P_{R}$. The transmit power of the source node $P_{S}$ is set to 30 dBm (dB in the following). The variances of RSI and the Gaussian noises are set to 1. The minimum data rate threshold is set to 3 bps/Hz. The results show that within the whole regime of the transmit power $P_{R}$, the proposed hybrid HD/FD relaying scheme outperforms the traditional hybrid HD/FD relaying scheme. Especially, in the high transmit power regime of $P_{R}$, the proposed hybrid HD/FD relaying scheme dramatically improves the system outage performance compared to the traditional HD/FD relaying scheme. The theoretical evaluations of both the traditional and the proposed HD/FD scheme are validated by the Monte Carlo (M-C in the figure) simulations.

In Fig. 4, we compared the system outage probability of the traditional HD/FD relaying scheme proposed in \cite{5961159} with that of the proposed HD/FD relaying scheme with the minimum data rate threshold $R_{0}$. The transmit power of $P_{S}$ and $P_{R}$ are set to 30 dB. The simulation results demonstrate the superiority of the proposed hybrid HD/FD relaying scheme over the traditional HD/FD relaying scheme. Especially in the low data threshold $R_{0}$ regime, the system outage performance is improved by two or three orders of magnitude. These insights have revealed that when all the antennas of the FD node is utilized within the HD mode, the system performance can be greatly improved by the incurred diversity gain.

\begin{figure}
  \centering
  % Requires \usepackage{graphicx}
  \includegraphics[width=3.6 in]{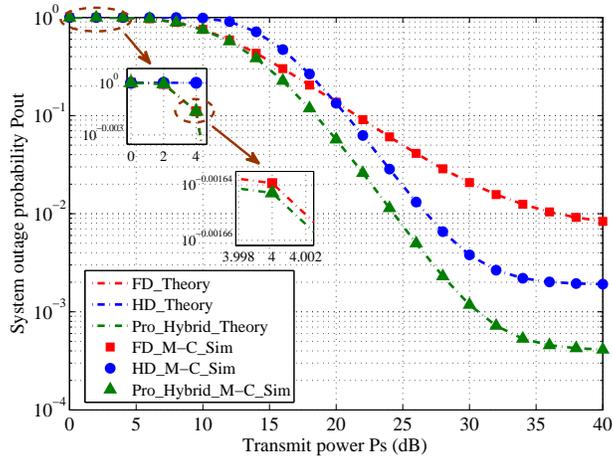}\\
  \vspace{0mm}
  \caption{The system outage probability of the proposed scheme vs. $P_S$.}\label{examp1}
\end{figure}

\begin{figure}
  \centering
  % Requires \usepackage{graphicx}
  \includegraphics[width=3.6 in]{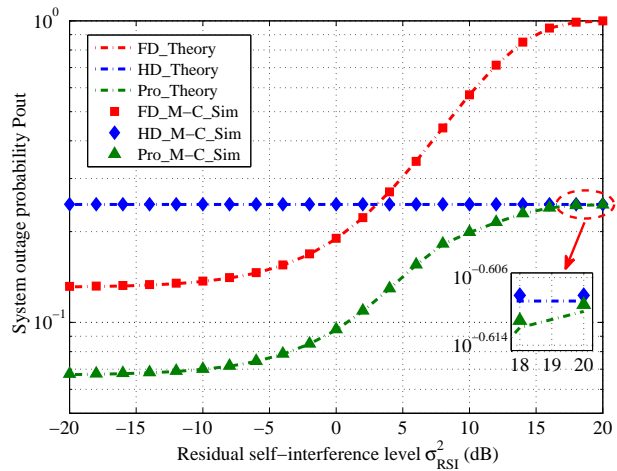}\\
  \vspace{0mm}
  \caption{The system outage probability of the proposed scheme vs. $\sigma_{RSI}^2$.}\label{examp4}
\end{figure}

In Fig. 5, we plot the system outage probabilities of the proposed hybrid HD/FD relaying scheme, the considered FD mode and the HD mode with the the transmit power $P_{S}$ . From the simulation results we can observe that in the low transmit power regime of $P_{S}$, the FD mode outperforms the HD mode. However, with the utilizing of the two antennas, the outage probabilities of the HD mode decrease faster than that of the FD mode with the increase of the transmit power $P_{S}$. In the high transmit power regime, the HD mode outperforms the FD mode. Nevertheless, the proposed hybrid HD/FD relaying scheme can always achieve better performance than the FD mode and the HD mode in the low transmit power and high transmit power regimes of $P_{S}$. %The results indicate that the proposed hybrid HD/FD relaying scheme could alleviate the effects of the residual self-interference as well as introduce the diversity gain.

In Fig. 6,  we plot the outage probabilities of the FD mode, the HD mode and the proposed hybrid HD/FD relaying scheme with the RSI. The results show that the RSI would severely limit the system outage performance of the FD mode whereas the outage probabilities of the HD mode are irrelevant to the RSI. In addition, it is noted that when the self-interference can be suppressed to a certain threshold, the FD mode is superior to the HD mode. However, the proposed hybrid HD/FD relaying scheme always achieves a better performance than the FD mode and HD mode within the whole RSI regime. This indicates that the proposed hybrid HD/FD relaying scheme can restrict the severe adverse effects of the RSI. %Furthermore, in the high level of RSI, the HD mode can serve as an upper bound of the system outage probabilities of the proposed hybrid HD/FD relaying scheme. %This indicate that within the high residual self-interference regime, the system will always work in the HD mode, thus, the outage probabilities of the HD mode can serves as an upper bound of that of the hybrid scheme when the self-interference incurred by the FD mode cannot be effectively eliminated.
%\vspace{-3mm}
\section{Conclusion}
In this paper, we proposed a new hybrid HD/FD relaying scheme, in which the two antennas of the FD relay node could be fully utilized when the system worked in the HD mode. In addition, we adopted the MRC/MRT technique to combine and transmit the signals to reap the diversity gain. The simulation results showed that the proposed hybrid HD/FD relaying scheme could dramatically improve the system outage performance compared to the traditional hybrid HD/FD relaying scheme. Moreover, the proposed hybrid HD/FD relaying scheme could more efficiently alleviate the adverse effects of the RSI incurred by the FD mode.

%\vspace{-1mm}

\bibliographystyle{IEEEtran}
\bibliography{reference}
\end{document}